\documentclass[12pt,a4paper]{article}
\usepackage{graphicx}
\usepackage{amsmath}
\usepackage{amssymb}
\usepackage{cite}
\usepackage[height=8.8in,width=6.45in]{geometry}
\usepackage{jheppub}
\usepackage{graphicx}
\usepackage{caption}
\usepackage{subcaption}

\numberwithin{equation}{section}

\begin{document}
\begin{titlepage}
\hfill CALT-TH 2018-020, IPMU18-0100\\
\vbox{
    \halign{#\hfil         \cr
           } 
      }  
\vspace*{15mm}
\begin{center}
{\Large \bf De Sitter Space and the Swampland}

\vspace*{15mm}

{\large Georges Obied,$^a$ Hirosi Ooguri,$^{b,c}$ Lev Spodyneiko,$^b$ Cumrun Vafa$^a$}
\vspace*{8mm}

${}^a$
Jefferson Physical Laboratory,
Harvard University, Cambridge, MA 02138, USA\\
\vspace*{0.2cm}
${}^b$ Walter Burke Institute for Theoretical Physics \\ California
Institute of Technology,
 Pasadena, CA 91125, USA\\
\vspace*{0.2cm}
${}^c$ Kavli Institute for the Physics and Mathematics of the Universe \\
University of Tokyo,
Kashiwa, 277-8583, Japan\\

\vspace*{0.7cm}

\end{center}
\begin{abstract}

It has been notoriously difficult to construct
a meta-stable de Sitter (dS) vacuum in string theory in a controlled approximation. This
suggests the possibility that meta-stable dS belongs to the swampland.
In this paper, we propose a swampland criterion
in the form of $|\nabla V|\geq\ c  \cdot V$ for a scalar potential $V$ of any consistent theory of
quantum gravity,
with a positive constant $c$.
In particular, this bound forbids dS vacua.
 The existence of this bound is motivated by the abundance of string theory constructions and no-go theorems which exhibit this behavior.  We also extend some of the well-known no-go theorems for the existence of dS vacua in string theory to more general accelerating universes and reinterpret the results in terms of restrictions on allowed scalar potentials.

\end{abstract}

\end{titlepage}

\vskip 1cm

\section{Introduction}

We live in a universe where the vacuum energy is positive.  This can be realized by having a scalar field potential
$V$ with a local minimum at a positive value, leading to a stable or meta-stable de Sitter (dS) vacuum.  However it could also be that the potential is positive but the scalar field is not at a minimum, as in quintessence models, as long as $|\nabla V|$ is sufficiently small and of the order of $V$ itself (for a review of quintessence models, see~\cite{Tsujikawa:2013fta}).
Most of the effort in the string theory community to realize our universe has been devoted to constructing dS vacua,
for example, in \cite{Maloney:2002rr,Kachru:2003aw,Balasubramanian:2005zx,Westphal:2006tn,Dong:2010pm,Rummel:2011cd,Blaback:2013fca,Cicoli:2013cha,Cicoli:2015ylx}.  Despite the heroic efforts in these attempts to construct such vacua, there has been a number of issues raised, for example, in
\cite{Maldacena:2000mw,Townsend:2003qv,Hertzberg:2007wc,Covi:2008ea,Caviezel:2008tf,Caviezel:2009tu,deCarlos:2009fq,Wrase:2010ew,Shiu:2011zt,Green:2011cn,Gautason:2012tb,Bena:2014jaa,Kutasov:2015eba,Quigley:2015jia,Dasgupta:2014pma,Junghans:2016abx,Junghans:2016uvg,Andriot:2016xvq,Moritz:2017xto,Sethi:2017phn,Andriot:2017jhf,Danielsson}.
It is thus fair to say that these scenarios have not yet been rigorously shown to be realized in string theory.

Given the difficulties in obtaining dS-like vacua in string theory, it is natural to contemplate the alternative
possibility where no dS, not even meta-stable dS, can exist in a consistent quantum theory of gravity (see~\cite{Brennan:2017rbf} as well as~\cite{Danielsson}
for raising this possibility).
One may be tempted to conjecture $|\nabla V| >A$ for some constant $A>0$, which
leads to such an exclusion. However,
supersymmetric vacua with flat directions provide counter-examples.
To avoid them, we may consider allowing $A$ to depend on scalar fields $\phi$ in such a way that $A(\phi) \leq 0$
for supersymmetric vacua. A simple form for such a dependence would be
$A(\phi)=c \cdot V(\phi)$ with positive constant $c$, since $V\leq 0$ in supersymmetric vacua.

With this motivation, we propose,
\begin{equation}
|\nabla V| \geq c\cdot V,
\label{bound}
\end{equation}
as a swampland criterion in any low energy theory of a consistent
quantum theory of gravity (for a recent review of other swampland criteria see \cite{Brennan:2017rbf}).  Here,
the norm $|\nabla V|$ of the potential gradient is defined using the metric on field space in the kinetic term of the scalar fields, and $c$ can depend on the macroscopic dimension of spacetime $d$.
The inequality means that the slope of $V$ cannot be too small when $V>0$.
We also propose that $c$ is of order 1 in Planck units.
If we take the Planck mass to infinity while keeping other variables fixed, the inequality becomes $|\nabla V| \geq 0$ and is trivially satisfied, as expected for a swampland criterion.
Though we have not been able to determine
the value of $c$, we will give an upper bound
based on a variety of top-down constructions from
string theory.

Let us discuss a couple of simple examples to illustrate the conjecture (\ref{bound}).
They are potential counter-examples, which will turn out to be consistent with the inequality.
Consider a supersymmetric vacuum with zero cosmological constant with a flat direction.
Since $|\nabla V|/V = 0/0$ is ill-defined, we need to deform the theory to test the inequality.
Since all parameters in string theory are dynamical,
any continuous deformation of the supersymmetric vacuum will involve a massive field,
$$V={1\over 2}m^2\phi^2.$$
If we are at $\phi \not= 0$,
$${|\nabla V|\over V}={2\over |\phi|}.$$
This can violate our conjecture when $\phi$ grows large.
However, the swampland criteria proposed in \cite{Ooguri:2016pdq}
claim that changing the value of $\phi$ beyond the Planck scale should
give rise to new light states, which would invalidate the simple potential in the above.
We can trust the potential above at most in a region where $|\phi| \lesssim 1$ in Planck units.
In this case, we see that $|\nabla V/ V|>\mathcal{O}(1)$, which is perfectly consistent with our conjecture.\footnote{This example
motivates the following weaker version of our conjecture, closer in spirit to \cite{Ooguri:2016pdq}:
As we get close to small $|\nabla V|/V$, a tower of light modes must appear with masses of the order of
 ${\rm exp}(-\tilde{c} \ V/|\nabla V|)$ in Planck units and invalidates the effective field theory.
 This may dovetail nicely with Vasiliev-type constructions of dS space, where there are infinitely many higher spin massless states. See~\cite{Didenko:2014dwa} for a review of Vasiliev theory.}
Another case to study, which also occurs commonly in string theory, is when $V$ crosses 0.
Consider a supersymmetric AdS solution and a massive field $\phi$.
Since $|\nabla V|$ is some non-zero constant when $V$ is crossing 0,
$|\nabla V|/V$ diverges.
Thus, the conjecture (\ref{bound}) passes some preliminary checks.

Note that the bound (\ref{bound}), while forbidding dS vacua, lends support to quintes\-sence models.
In quintessence models, $|\nabla V|$ typically has to be of the order of $V$, which is small
in the current universe when measured in Planck units. However, the inequality (\ref{bound})
suggests that such a small value of $|\nabla V|$ is natural since it would be close to the universal
swampland bound.   Moreover, a small enough slope will lead to accelerating expansion.  For example, in four dimensions, if $V$ is positive and we consider a single scalar field with
\begin{equation}
    \frac{|\partial V|}{V} = \lambda < \sqrt{6},
\end{equation}
constant, the $w$-parameter relating
the energy density $\rho$ and the pressure $P$ as $P = w \rho$ approaches asymptotically to,
\begin{equation}
    w = -1 + \frac{\lambda^2}{3}.
    \label{wparameter}
\end{equation}
The accelerating expansion of the universe requires $w < -1/3$, namely, $\lambda < \sqrt{2}$.
This would be compatible with the inequality (\ref{bound}) provided $c < \sqrt{2}$ for $d=4$.  It is shown in \cite{NEW} that based on current experimental data $c\lesssim 0.6$ in 4 dimensions.
As we will see in this paper, the experimental bound on $c$ is tantalizingly close to and only slightly less
than the values we obtain based on various restrictive assumptions in string theory.

The main purpose of this paper is to review and examine string-theoretical constructions to motivate and test
the conjecture (\ref{bound}).   We emphasize that all bounds we present
in this paper are only meant to serve as upper bounds to the true value of $c$ because the assumptions we make in deriving them are too restrictive and can be relaxed. In particular, the bounds we obtain will hold only for the subset of string vacua for which those restrictive assumptions are obeyed.  Nevertheless, it is important to note that the bounds that we obtain under a variety of assumptions all lead to $c\sim \mathcal{O}(1)$.
 In Section 2, we review examples from the literature, and generalize and extend them in various ways.
 We will take into account quantum and stringy effects as well as classical effects.  This includes scaling arguments in M theory and also Type II and heterotic string constructions. We also study consequences of the strong energy condition (SEC) and the null energy condition (NEC), though both of them are known to be too restrictive in string theory.
 In section 3, we point out another issue in constructing accelerating universes in string theory.
In  \cite{Farhi:1986ty},  Farhi and Guth showed that it is not possible to realize a de Sitter space
in an asymptotically flat space from a smooth initial condition. We extend their result in four dimensions to
an accelerating universe with $|\nabla V|/V < \sqrt{2}$.
 In Appendix A we discuss the Type II no-go theorems and in Appendix B we discuss further no-go theorems in certain classes of models.  In Appendix C, we discuss the consequences of assuming SEC and NEC on $c$.

\section{Examples}

In this section, we study examples from M theory and string theory constructions.
We first consider M theory compactified on a large manifold, where the supergravity approximation is applicable, and study
the potential it generates in lower dimensions, where the infinitely many scalar fields come from all possible metric and flux variations on the compactified manifold.  We use scaling arguments in this case to get a bound on $c$.   Another example we will consider is the $O(16)\times O(16)$ heterotic string. This is a genuinely non-supersymmetric string vacuum in $10$ dimensions, which has a positive vacuum energy~\cite{AlvarezGaume:1986jb} but no tachyons.  We will show this leads to a value of $c$ of order 1.
We then consider a number of additional examples involving Type II compactifications with fluxes and orientifold planes. Some of these examples are from the existing string literature and others are additional examples we study. We will end this section by discussing bounds for $c$ obtained from SEC and NEC.

While we will not be able to close possible loopholes in all conceivable constructions,
the diversity of examples and the simplicity of the arguments used in these cases to derive the bound (\ref{bound}) lends support to our conjecture.

\subsection{M Theory Compactifications}
Let us start with the $11$-dimensional supergravity with the bosonic part of the action,
\begin{equation}
  2\kappa_{11}^2 S = \int d^{11}x\sqrt{-g^{(11)}}\Big(\mathcal{R}-\frac{1}{2}|G|^2\Big),
  \label{mtheoryaction}
\end{equation}
and consider compactifications to $d$ dimensions on an arbitrary manifold with no singularities,
 volume $V_{11-d} = e^{(11-d)\hat{\rho}}$, and curvature much less than Planck scale
so that  we can trust the supergravity approximation. We consider the class of metrics that take the form\footnote{
Though this ansatz could be generalized by allowing
a $y$-dependent warp factor multiplying the metric $g_{\mu\nu}(x) dx^\mu dx^\nu$ of the macroscopic dimensions,
we can set it to be $1$ without loss of generality for the purpose of deriving the gradient of the scalar potential $V$,
as we will explain in Section 2.4 and Appendix C.}\textsuperscript{,}\footnote{Note that in our notation moduli fields with and without a hat are related such that $\ln{\rho}\propto \hat{\rho}$. This is consistent with the notation in~\cite{Hertzberg:2007wc} and with our notation used in later sections.}
\begin{equation}
ds^2 = dx_d^2 + e^{2\hat{\rho}(x)}dy_{11-d}^2 = g_{\mu\nu}(x) dx^\mu dx^\nu + e^{2\hat{\rho}(x)} \gamma_{ab}(y) dy^a dy^b.
\label{Mansatz}
\end{equation}
Maldacena and Nu\~{n}ez
\cite{Maldacena:2000mw} have proved that one cannot obtain a dS vacuum in such a setup.
In this section, we extend their no-go theorem in the context of M theory to show that, not only we cannot get dS,
where $\nabla V=0 $ and $V>0$, but we cannot even come close to $\nabla V=0$.

In the effective theory in $d$ dimensions, the metric moduli and flux data of the internal geometry play the role of scalar fields. Even though there are many (and in fact infinitely many) such scalar fields controlling the details of the compactification, we will concentrate only on the overall volume modulus $\hat{\rho}$, which for sure is one of the degrees of freedom of the metric.
If we can show that the gradient of the potential along this direction is large, then that is sufficient to show that
$|\nabla V|$, which probes the gradient in all directions, is large.
There will be two contributions to the effective potential setting the volume modulus $\hat{\rho}$:
(i) from the flux term and (ii) from the curvature term of the internal manifold. We discuss each contribution separately.

First, we have the term from the Ricci scalar of the compact dimensions. To an observer in the macroscopic
dimensions, this Ricci curvature will manifest itself only as an integrated and averaged quantity given by $e^{-2\hat{\rho}}\mathcal{\bar{R}}_{11-d}$, where $\mathcal{\bar{R}}_{11-d}$ is the average Ricci scalar when the volume of the internal manifold is set to 1. The contribution from this term to the effective potential will have an overall $\hat{\rho}$-dependent factor given by the product $e^{-2\hat{\rho}} \cdot e^{(11-d)\hat{\rho}} \cdot e^{\frac{-d(11-d)}{d-2}\hat{\rho}} \propto e^{-18\frac{\hat{\rho}}{d-2}}$, where the second and third factors come from the metric determinant and the Weyl scaling respectively. Moreover, the sign of this contribution is opposite that of the integrated curvature. This means
that
negatively curved internal manifolds tend to decompactify and vice versa. Second, we have the contribution from $G$-flux which is always positive and carries an overall $\hat{\rho}$-dependent factor that is given by $e^{(11-d)\hat{\rho}}\cdot e^{-8\hat{\rho}}\cdot e^{\frac{-d(11-d)}{d-2}\hat{\rho}} \propto e^{-6\frac{d+1}{d-2}\hat{\rho}}$, where the exponentials come from the metric determinant, the 4 inverse metric factors, and the Weyl rescaling respectively.  To obtain a canonically normalized scalar field, with the kinetic term ${1\over 2} (\nabla \hat{\rho} )^2$, we need to scale $\hat{\rho} \rightarrow \sqrt{\frac{9(11-d)}{d-2}} \hat{\rho}$.  This leads to the effective potential
\begin{equation}
  V= V_\mathcal{R}\ e^{-\lambda_1 \hat{\rho}} + V_G\ e^{-\lambda_2 \hat{\rho}},
\end{equation}
\noindent where $\lambda_1 \equiv \frac{6}{\sqrt{(d-2)(11-d)}}$, $\lambda_2 \equiv \frac{2(d+1)}{\sqrt{(d-2)(11-d)}}$,
$V_{\mathcal R}$ is proportional to the average of the Ricci scalar curvature, and $V_G$ (which is always positive) is proportional to the average of the G-flux term in the action.
If the manifold is on the average negatively curved and $V_{\mathcal R} \geq 0$ then $|\partial_{\hat{\rho}} V|/V$ is bounded below by the smaller of $\lambda_1$ or $\lambda_2$, namely, $\lambda_1$ for $d \geq 2$.

One can also consider cases when the average of the scalar curvature is positive, with $V_\mathcal{R} <0$. In this case, whenever $V < 0$, our bound is trivially satisfied. Moreover, for $V> 0$, it is easy to see that $|\partial_{\hat{\rho}} V|/V > \lambda_2$ by noting that it must diverge at $V = 0$, approaches $\lambda_2$ as $\hat{\rho}\rightarrow -\infty$ and that it is monotonic.

Though our argument above does not rely on supersymmetry, the bound on $|\partial_{\hat{\rho}} V|/V$ also manifests itself in well-known supersymmetric compactifications such as the Freund-Rubin solutions of the
$11$-dimensional supergravity giving rise to $AdS_4 \times S^7$ or $AdS_7 \times S^4$ spacetimes.
The metric on the large dimensions is taken to be that of $AdS$ and that on the compact dimensions is taken to be that of a sphere. The preferred radius is then set by the effective potential due to a balance between contributions from the positive flux term, which dominates at small radius, and the negative curvature term, which dominates at large radius. The regime where $V>0$ is of special interest. In this region of the field range, the flux term must dominate since its contribution is positive, and we again have $|\partial_{\hat{\rho}} V| > \lambda_2 V$ by considering limits similar to above.

Finally, the bounds on $|\nabla V|/V$ we have found in the examples above can be saturated in string theory constructions. For instance, the bound in the case of $AdS_4 \times S^7$ is saturated in the limit of small $S^7$ volume.  One may worry that this is not strictly trustable because the curvature may be too large.  However, by choosing the flux to be large enough one can see that $|\nabla V|/V$ approaches the above bound before entering a Planckian curvature regime.
Note that any bound presented for this case can be viewed as an upper bound on the constant $c$
since the latter should be considered as the infimum of all $|\nabla V|/V$ over all (reliable) string constructions.

To conclude, we find that compactification of the $11$-dimensional supergravity to $d$ dimensions gives the bound,
\begin{equation}
  \frac{|\nabla V|}{V} \geq \frac{6}{\sqrt{(d-2)(11-d)}}.
  \label{mtheorybound}
 \end{equation}
 In particular for $d=4$, this yields $6/\sqrt{14}\sim 1.6$ which gives an upper bound on $c$ in our conjecture because it is actually realized by the $AdS_4 \times S^7$ example.
  As we will discuss in subsection 2.4, this inequality is also a consequence of the fact that
 the $11$-dimensional action (\ref{mtheoryaction}) satisfies the strong energy
 condition (SEC). Since the SEC is often violated in string theory, let us discuss more stringy examples in the following
 two subsections and see whether we have similar bounds in string theory.

\subsection{$O(16)\times O(16)$ Heterotic String}

To study effects beyond the supergravity approximation, let us examine
compactifications of the non-supersymmetric heterotic string
 \cite{Dixon:1986iz,AlvarezGaume:1986jb} constructed by twisting the $E_8\times E_8$ theory,
 which has a $10$-dimensional cosmological constant $\Lambda>0$ in its low-energy effective action due to string one-loop effects \cite{AlvarezGaume:1986jb}. If one considers the effective $4$-dimensional
potential as a function of a characteristic compactification size $R$ for a fixed dilaton background,
one has $V_{eff}(R\rightarrow \infty) \rightarrow \Lambda$. This suggests the possibility of finding a dS vacuum,
perhaps by creating a minimum in the effective potential at some finite radius $R$. We  are going to
show that this is not possible.

The non-supersymmetric heterotic string has the effective action at weak coupling given by:
\begin{equation}
  S = \int d^{10}x \sqrt{-g^{(10)}}\left[ \frac{1}{2\kappa^2} e^{-2\phi}\left(\mathcal{R} + 4(\partial\phi)^2  \right) - \Lambda
  \right],
\end{equation}
\noindent where $\phi$ is the dilaton and we work in a regime where curvature is small compared to the Planck scale. Consider first the 10-dimensional value of $|\nabla V|/V$. Weyl rescaling to Einstein frame ($g_{\mu\nu} \rightarrow e^{-\phi/4}g_{\mu\nu}$) and defining the canonical field $\hat{\tau} \equiv \frac{1}{\sqrt{2}} \phi$ gives the effective potential:
\begin{equation}
  V_{eff} = V_\Lambda e^{-5{\hat \tau}/\sqrt{2}},
\end{equation}
\noindent where $V_\Lambda > 0$ is a constant. We thus have $|\nabla V|/V = 5/\sqrt{2} \approx 3.5$.  Since this case is realized (at least in the extreme weak coupling limit ${\hat \tau} \gg 1$ where we can trust the string one loop computation) we learn that for $d=10$, the value of $c$ in our conjecture satisfies\footnote{Note that we have ignored stringy modes in this discussion.  It would be interesting to check whether we can ignore these in an effective action for the rolling scalar field, as it could lead to their production.} $c\leq 3.5$.

Consider now the case where we compactify to $d<10$ dimensions on an arbitrary manifold with volume $e^{(10-d)\hat{\rho}}$ and for simplicity let us not turn on any additional fields. This gives a two-dimensional scalar field space parameterized by ($\hat{\rho},\phi)$. The spacetime potential on this field space receives contributions from the $(10-d)$ dimensional Ricci scalar averaged over the internal manifold and the cosmological constant term. The former provides an overall $(\hat{\rho},\phi)$-dependent factor given by $e^{-2\hat{\rho}}\cdot e^{-2\phi} \cdot e^{(10-d)\hat{\rho}} \cdot e^{-\frac{d}{d-2}[(10-d)\hat{\rho} - 2\phi]} \propto e^{-\frac{4}{d-2}(4\hat{\rho} - \phi)}$, where the last two factor are from the metric determinant and Weyl scaling respectively. The latter term gives a contribution of the form $e^{(10-d)\hat{\rho}}\cdot e^{-\frac{d}{d-2}[(10-d)\hat{\rho} - 2\phi]} \propto e^{-\frac{2}{d-2}[(10-d)\hat{\rho} - d\phi]}$.

Defining the canonical field $\hat{\tau} \equiv \frac{2}{\sqrt{d-2}}(\phi - \frac{10-d}{2}\hat{\rho})$ and rescaling $\hat{\rho} \rightarrow \sqrt{10-d} ~\hat{\rho}$ gives the effective potential,
\begin{equation}
  \label{eq:hetpot}
  V_{eff} = V_\mathcal{R} e^{-\frac{2}{\sqrt{10-d}}\hat{\rho}+\frac{2}{\sqrt{d-2}}\hat{\tau}} + V_\Lambda e^{\sqrt{10-d}\hat{\rho} +\frac{d}{\sqrt{d-2}}\hat{\tau}},
\end{equation}
In this case, assuming average negatively curved space leading to positive $V_\mathcal{R}$, the quantity $|\nabla V|/V$ is bounded below by the smaller of $2\sqrt{(3d-5)/(d-2)}$ and \\ $4\sqrt{2}/\sqrt{(10-d)(d-2)}$, which is the latter for $3\leq d\leq 9$. For $d = 4$, we have $|\nabla V|/V \approx 1.6$. The conclusion for positively curved internal manifolds on the average, parallels that of the M theory case.

Unlike the M theory examples discussed in the previous subsection, where the inequality (\ref{bound}) can
be regarded as a consequence of the SEC in 11 dimensions (as we will explain in subsection 2.4),
the $O(16) \times O(16)$ example is not tied to the SEC. In fact, as previously discussed, it has a positive potential for a scalar in
$10$ dimensions, violating the SEC. This example demonstrates that the breaking of the SEC,
in particular the positive cosmological constant in higher dimensions, does not guarantee stable or meta-stable dS minima
after compactification and that our conjecture still continues to hold in such a case.
It is therefore not correct to claim that, because string theory (as opposed to supergravity theories) contains ingredients
that violate the SEC, it should necessarily be able to accommodate dS vacua.

It is interesting to note that for the usual supersymmetric heterotic string, statements about the lack of classical dS vacua can be made in the perturbative regime of the heterotic string by relying on world-sheet arguments~\cite{Kutasov:2015eba}.  This in particular includes all order $\alpha'$ stringy corrections to the Einstein equations.

\subsection{No-Go Theorems in Type II Supergravity}

There have been several results in the literature discussing no-go theorems on dS vacua in string theory constructions,
with restrictions on ingredients used in string theory, typically specific combinations of fluxes, D-branes, orientifolds, etc. For instance,~\cite{Hertzberg:2007wc} presented a general no-go theorem for dS vacua from Calabi-Yau
compactifications with $O6/D6$ and general NS-NS and R-R fluxes.  The argument also provides a bound $|\nabla V|/V > \sqrt{54/13}\sim 2$ in the volume-dilaton plane in four dimensions.  In
\cite{Wrase:2010ew}, constructions were considered that have orientifold sources, but without D-branes. We have been able
 to generalize these results
to include both orientifold and D-brane sources in diverse dimensions, and to compactifications on manifolds that are not Ricci-flat.

In order to generalize the existing results, we consider the effective potential given by\footnote{Note that, as before, $\hat{\rho}\propto \ln{\rho}$ and $\hat{\tau}\propto \ln{\tau}$.},
\begin{equation}
    V_{eff}= \frac{A_\mathcal{R}}{\tau^2 \rho} - \frac{A_O}{\tau^3 \rho^{(6-q)/2}}+\sum_i \frac{A_i}{\tau^{\alpha_i} \rho^{\beta_i}} ,
    \label{effectivepotential}
\end{equation}
\noindent where the first term comes from the curvature term, the second one from O$q$ planes, and the last sum is over contributions from NS-NS fluxes, R-R fluxes, and D-branes. We define the derivative with an arbitrary coefficient $\mathcal{D}(a) \equiv -a\tau\partial_\tau - \rho\partial_\rho$
and $C_i(a) = \mathcal{D}(a)V_i/V_i$ is defined for each term $V_i$ in the effective potential (\ref{effectivepotential})
 as in~\cite{Hertzberg:2007wc}. A no-go theorem can be found when $C_i^{(-)} \leq \min_j C_j^{(+)}$ for all $i$, where $C_i^{(\pm)}$ denotes the constant corresponding to a positive/negative contribution to the potential. This leads to a set of inequalities that the coefficient $a$ must satisfy. Whenever these hold for a single value of $a$, one can exclude the existence of a dS minimum and derive a corresponding limit on $|\nabla V|/V$.  See Appendix~\ref{app:typeIIextensions} for
 detailed analysis on this.  Our results are summarized in Table~\ref{tab:bounds}.  It shows the bounds on constructions with
  D$q$ branes and O$q$ planes of the same dimension, as well as arbitrary NS-NS and $F_r$ R-R fluxes (with the exceptions noted in the table).

\begin{table}
  \centering
\begin{tabular}{|l||l|c||l|c|}
  \hline
  $q$ & No-Go (positive or zero $\langle \mathcal{R}\rangle $) & $c^2_\star$ & No-Go (negative $\langle \mathcal{R}\rangle $) & $c^2_\star$\\
  \hline
  3 & Yes w/o $F_1$ RR flux & 6     & Indeterminate                  & -     \\
  4 & Yes w/o $F_0$ RR flux & 98/19 & Indeterminate                 & -     \\
  5 & Yes                   & 32/7  & Yes w/o $F_1$ RR flux & 2     \\
  6 & Yes                   & 54/13 & Yes w/o $F_0$ RR flux & 18/7  \\
  7 & Yes                  & 50/12 & Yes                   & 8/3   \\
  8 & Yes                   & 242/67& Yes                   & 50/19 \\
  9 & Yes                   & 24/7  & Yes                   & 18/7  \\
  \hline
\end{tabular}
\caption{\footnotesize Constrains on $|\nabla V|/V$ in Type IIA/B compactifications to 4
dimensions with arbitrary R-R and NS-NS flux (unless otherwise noted) and O$q$-planes and D$q$-branes with fixed $q$. The constant $c_\star$ in each entry is a lower bound on $|\nabla V|/V$.}
\label{tab:bounds}
\end{table}

The constant $c_\star$ shown in the table is not necessarily
the most stringent lower bound on $|\nabla V|/V$.
This is because compactifications typically have more moduli than those
we considered in deriving these bounds and
hence there might even be steeper directions in moduli space.
Moreover, the nature of the general scaling argument in this section makes it difficult to judge
whether these bounds are actually realized in string constructions.
Since all the values of $c_\star$ in the table are greater than $\sqrt{2}$, which means $w > -1/3$ by
(\ref{wparameter}), none of these examples give rise to an asymptotically accelerating expansion in four dimensions.

There are a couple of entries in the Table~\ref{tab:bounds} noted as ``Indeterminate.''
In these cases, we have not been able to exclude a point with $\nabla V = 0$ in the two-dimensional
subspace of $(\tau, \rho)$. However, we note that arguments made in \cite{Andriot:2016xvq} using Bianchi identities exclude the case $q=3$. Moreover, if one does not allow D-brane sources,
it was shown in \cite{Wrase:2010ew} that for $q=4$ we get a bound $c_\star=\sqrt{2/3}\sim 0.8$.  At any rate, just because we cannot determine a bound by this simple scaling argument does not mean that dS vacua exist in these cases since
$\nabla V$ may be non-zero in other directions in the moduli space.  In Appendix~\ref{app:instability},
we give examples in which we can estimate contributions from other directions and find  unstable directions expected by our
conjecture.

Similar to the M theory case above, one can consider well-known supersymmetric examples such as Type IIB on $AdS_5\times S^5$ with five-form flux threading the sphere. An effective potential can be defined on the volume-dilaton plane and takes the form:
\begin{equation}
  V_{eff} = -V_\mathcal{R} e^{-\frac{2}{\sqrt{5}}\hat{\rho}+\frac{2}{\sqrt{3}}\hat{\tau}} + V_F e^{-\frac{5}{\sqrt{5}}\hat{\rho} +\frac{5}{\sqrt{3}}\hat{\tau}}.
\end{equation}
\noindent Note that the curvature term is identical to Eq.~\ref{eq:hetpot} with $d=5$ due to similarity in the actions. Additionally, the flux term is identical to the cosmological constant term up to a factor of $e^{-10\hat{\rho}}$ from 5 factors of the inverse metric. A canonical kinetic term for $\hat{\rho}$ is defined by scaling $\hat{\rho} \rightarrow \sqrt{10-d}~\hat{\rho}$ and altogether this gives the above effective potential. This implies a bound $|\nabla V|/V \geq 2\sqrt{10/3}$, which is realized in the region of moduli space where the contribution from the flux dominates the effective potential corresponding to the small volume limit of $S^5$ which one may again hope to be trustable given the high degree of supersymmetry.  Since this value is actually realized in one example, it shows that for $d=5$ the actual $c\leq 2\sqrt{10/3} \sim 3.7$.

\subsection{Consequences of Energy Conditions}

The bound on $|\nabla V|/V$ for compactifications of
the $11$-dimensional supergravity
derived in section 2.1 is in fact a consequence
of SEC in $11$ dimensions.
Suppose more generally we start with a theory in $D$ dimensions
and compactify it on a $(D-d)$-dimensional space with a metric
$g_{mn}$ and a warp factor $\Omega$ as,
\begin{equation}
ds^2= \Omega(y,t)^2\left( -dt^2 + a(t)^2 d\vec x^2_{d-1}\right) + g_{mn}(y, t) dy^m dy^n.
  \label{SECansatz}
\end{equation}
In Appendix~\ref{app:SECNEC}, we derive the
bound on the gradient of the effective potential $V$ in
$d$ dimensions,
\begin{equation}
   \frac{|\nabla V|}{V} \geq \lambda_{SEC} \equiv 2\sqrt{\frac{D-2}{(D-d)(d-2)}},
\label{SECbound}
 \end{equation}
 assuming that the $D$-dimensional theory satisfies SEC. This
 result is independent of the choice of the internal metric $g_{mn}$.
 Note that the M theory action (\ref{mtheoryaction}) in 11 dimensions satisfies SEC. Indeed,
 the bound (\ref{mtheorybound}) on M theory compactifications we found there
 is reproduced by setting $D=11$, and therefore is simply a consequence of SEC.

However, we need to take this result with a grain of salt since SEC can easily be violated in string theory.
We should also note that an accelerated expansion of the universe in $d$ macroscopic dimensions
requires $(d-3)\rho +(d-1)p<0$.
 For a constant exponential potential $V=V_0\ {\rm exp} (-\lambda \phi)$,
 a solution approaches $p=w\rho$ with $w=-1+{1\over 2}{d-2\over d-1} \lambda^2$,
 leading to,
\begin{equation}
  \frac{|\nabla V|}{V} < \lambda_{accelerate} \equiv \sqrt{\frac{4}{d-2}}.
  \label{accelerationbound}
  \end{equation}
  This is incompatible with (\ref{SECbound}) $d> 2$. Clearly, SEC is too strong an assumption to make.

We may also consider NEC, which is satisfied
by a broader range of
theories, though it can be violated quantum mechanically.
To analyze consequences of NEC, it is convenient to parametrize the metric as,
\begin{equation}
ds^2= \Omega(y,t)^2\left( -dt^2 + a(t)^2 d\vec x^2_{d-1}\right) +\Omega(y,t)^{-\frac{2d}{D-d-2}} g_{mn}(y,t) dy^m dy^n.
  \label{NECansatz}
\end{equation}
The power of $\Omega$ in front of $g_{mn} dy^m dy^n$ is a matter of convention since we can always redefine $g_{mn}$
to absorb this factor. However, unlike the case of SEC, we need to make an additional assumption on the internal metric $g_{mn}$,
that the average of its scalar curvature
is either zero or negative, in order to derive a bound on $V$. Under this assumption, we find,
\begin{equation}
    \frac{|\nabla V|}{V} \geq \lambda_{NEC} \equiv 2\sqrt{\frac{D-d}{(D-2)(d-2)}}.
   \label{NECbound}
 \end{equation}
This generalizes the result of
\cite{Steinhardt:2008nk}
that NEC in $D$ dimensions together with a similar assumption on the internal metric $g_{mn}$
excludes a dS solution in macroscopic dimensions.
This NEC bound is compatible with an accelerated expansion of the universe, namely,
\begin{equation}
     \lambda_{NEC} < \lambda_{accelerate} < \lambda_{SEC},
 \end{equation}
 for $d >2$.  However, this is still too strong an assumption.  In particular experimental bounds give a value for $c\lesssim 0.6$ \cite{NEW} whereas the NEC bound (\ref{NECbound}) -- under the assumptions that the supergravity description is
 valid and the internal space has either zero or negative average scalar curvature -- would give the bound $\lambda_{NEC}=\sqrt{3/2}\sim 1.2$ for $d=4$ and $D=10$.

 It would be interesting to find out if a bound on $|\nabla V|/V$ can be obtained with the averaged null energy condition,
 which is weaker and is known to be satisfied by any unitary quantum field theory
 \cite{Hartman:2016lgu, Balakrishnan:2017bjg}.

\section{Accelerating Universe in a Laboratory}

 Suppose a scalar potential $V$ is such that it allows an asymptotically flat space.
 We will show that, even if there is a region in the field space in which $V > 0$ and,
\begin{equation}\label{censorship}
|\nabla V|/V < \lambda_{accelerate}  \equiv \sqrt{\frac{4}{d-2}},
\end{equation}
such a region cannot be reached by a classical time evolution starting with
 a smooth initial condition in an asymptotically flat space.
This can be regarded as generalization of the result \cite{Farhi:1986ty} by Farhi and Guth, from dS to small gradient potentials.

Let us start with a theory in which the vacuum is at $V=0$ and consider a massive field in the
theory whose excitation gives $V(\phi) > 0$.
In~\cite{Farhi:1986ty}, it was shown that one cannot start with an asymptotically flat space at $V=0$
 and create a region which is in the meta-stable dS minimum while avoiding initial singularities.
 This is because such a configuration creates an anti-trapped surface inside the dS region,
and Penrose theorem shows that there should have been an initial singularity. This result
 can be sharpened by considering not only potentials with a dS minimum, but also nearly-flat potentials since the same issue arises in these scenarios.

We assume an effective potential that has a minimum at $V(\phi_0) = 0$ and also has some field range over
which (\ref{censorship}) is satisfied.  One can imagine starting with $\phi=\phi_0$ and
 locally exciting the field in order to reach (\ref{censorship}).
 Such a region would drive
 an accelerating expansion of spacetime.
 We should place the domain wall separating the region (\ref{censorship}) and $\phi= \phi_0$
 outside of the cosmic event horizon of the accelerating region so that an observer in
the accelerating universe does not see the wall (this is the same condition required by  \cite{Farhi:1986ty}).

In the accelerating universe, there is an anti-trapped surface, which is a compact surface where light rays
directed inwards and outwards are both diverging. Since we have chosen
the domain wall to be outside of the cosmological event horizon, the anti-trapped surface is not affected by it.
Since all configurations of a scalar field minimally coupled to gravity
obeys NEC, the assumptions of the Penrose theorem are also satisfied and an initial singularity is inevitable.
It makes it impossible to probe the flat potential starting with a smooth initial condition.
Note that the same conclusion is not reached for a steeper potential.

The presence of the initial singularity itself may not be an issue. However, for a meta-stable dS realized
in an asymptotically AdS space, the initial singularity has led to various puzzles on its holographic meaning
\cite{Freivogel:2005qh, Fischetti:2014uxa}. It is straightforward to generalize these observations to
accelerating universes in an asymptotically anti-de Sitter space, posing challenges to their holographic interpretations.

\section*{Acknowledgments}

We would like to thank P.~Agrawal, G.~Horowitz, S.~Kachru, D.~Marolf, M.~Sasaki, S.~Sethi, P.~Steinhardt and X.~Yin for discussions.  We would also like to thank SCGP Summer Workshop 2017 for hospitality during part of this work.
The research of HO is supported in part by
U.S.\ Department of Energy grant DE-SC0011632,
by the World Premier International Research Center Initiative,
MEXT, Japan,
by JSPS Grant-in-Aid for Scientific Research C-26400240,
and by JSPS Grant-in-Aid for Scientific Research on Innovative Areas
15H05895.
HO also thanks the hospitality of
the Aspen Center for Physics, which is supported by
the National Science Foundation grant PHY-1607611.  The research of CV is supported in part by the NSF grant
PHY-1067976.

\appendix
\section{Extensions of no-go theorems in Type II supergravity}
\label{app:typeIIextensions}
We start with the effective potential\footnote{We are assuming $A_O>0$, i.e. orientifolds with negative tension.   If $A_O\leq 0$ we can get even a stronger bound.}
\begin{equation}
    V_{eff}= \sum_i \frac{A_i}{\tau^{\alpha_i} \rho^{\beta_i}} +\frac{A_\mathcal{R}}{\tau^2 \rho} - \frac{A_O}{\tau^3 \rho^{(6-q)/2}},
\end{equation}
and define the derivative $\mathcal{D}(a) \equiv -a\tau\partial_\tau - \rho\partial_\rho$ such that it has the following action on the different contributions
\begin{align}
  \mathcal{D}V_i &= (a\alpha_i + \beta_i)V_i \equiv C_i V_i,\\
  \mathcal{D}V_\mathcal{R} &= (2a+1)V_\mathcal{R} \equiv C_\mathcal{R} V_\mathcal{R},\\
  \mathcal{D}V_{Oq} &= (3a + \frac{6-q}{2})V_O \equiv C_O V_i.
\end{align}

We wish to choose a value for $a$ such that $\mathcal{D}V = K V + \text{(positive)}$ where $K>0$ is a constant. As such we can group the $C_i$ into two groups, $C_i^{(+)}$ and $C_i^{(-)}$, according to whether the contribution to the potential is positive or negative. A no-go theorem then exists when $C_i^{(-)} \leq \min_j C_j^{(+)}$ for all $i$.

For a flat curved internal manifold, with $F_r$ RR flux, O$q$-planes and D$p$-branes, the above reduces to the following set of inequalities
\begin{align}
  2a + 3 &> 0 \label{eq1_1},\\
  4a + (r-3) &> 0 \label{eq2_1},\\
  3a + \frac{6-p}{2} &> 0 \label{eq3_1},\\
  2a - q  &\le 0 \label{eq4_1},\\
  a + (r-3) - \frac{6-q}{2} &\ge 0 \label{eq5_1},\\
  p-q &\le 0 \label{eq6_1}.
\end{align}
For a negatively curved internal manifold, these are augmented by
\begin{align}
  2a + 1 &> 0 \label{eq7_1},\\
  a + \frac{4-q}{2} &\le 0 \label{eq8_1}.
\end{align}
Similarly, for a positively curved internal manifold, these are augmented by
\begin{align}
  2a + (r-4) &\ge 0 \label{eq7},\\
  a + \frac{4-p}{2}&\ge 0\label{eq8}.
\end{align}
We find all combinations of 3-tuples $(p, q, r)$ with $p=q$ for which $\exists a>0$ that satisfies the above inequalities. For these cases, one has the result $(|\nabla V|/V)^2 \geq c^2_\star$ which we report in Table~\ref{tab:bounds}.  This is obtained
exactly as in the method used in \cite{Hertzberg:2007wc,Wrase:2010ew}.

\section{Unstable Directions in Moduli Space}
\label{app:instability}

We now consider an example where the no-go theorems of the previous section seem to fail and show that the existence of dS vacua is still unclear if one takes into account additional directions in moduli space~\cite{Caviezel:2008tf}. The example is that of Type IIA supergravity compactified on a 6D manifold with negative curvature. In order to avoid the above no-go theorems, we include an O$4$-plane wrapping a homologically trivial cycle and $F_{2,4}$ R-R fluxes. We assume a product structure for the internal manifold so that we can rescale the directions parallel vs. perpendicular to the O-plane independently.

To begin with, we consider scaling in the volume-dilaton plane. Defining the moduli
\begin{equation}
  \rho = (\rm Vol_6)^{1/3}, \quad \tau = e^{-\phi}\sqrt{Vol_6},
\end{equation}
\noindent we work out the $(\rho,\tau)$-dependence of the contributions to the effective 4D potential. For instance, the term from the internal curvature scales as $\tau^2 \rho^{-1} \cdot \tau^{-4}$ where the last factor is from Weyl rescaling in the 4 non-compact dimensions. Similarly, considering other contributions leads to a potential of the form
\begin{equation}
V_{eff} = \frac{A_\mathcal{R}} { \tau ^ { 2} \rho }
    - \frac { A_O } { \tau ^ { 3} \rho }
    + \frac { A_2 \rho } { \tau ^ { 4} }
    + \frac { A_4 } { \tau ^ { 4} \rho },
\end{equation}
\noindent where the coefficients $A_i$ generally depend on other moduli we are neglecting. Na\"{i}vely, the above potential can have a de Sitter minimum if one can find a string theory construction where the coefficients satisfy the condition
\begin{equation}
  192 \leq \frac{(7A_O)^2}{A_\mathcal{R} A_4} \leq 196.
\end{equation}
We show that allowing the internal manifold to scale independently along the directions transverse to the orientifold plane removes the de Sitter solution. In this case, the moduli with diagonal kinetic terms are given by:
\begin{equation}
  \pi = V_1^{2/k_1};\quad \sigma = V_2^{2/k_2};\quad\tau = e^{-\phi}\sqrt{V_1 V_2},
\end{equation}
\noindent where $V_{i}$ are the volumes of the subspaces parallel and transverse to the O-plane and $k_i$ are their dimensions. In the following, we consider an example based on a compactification with an O$4$-plane wrapping a homologically trivial cycle and arbitrary $F_{2,4}$ R-R flux. We allow the rescaling of directions parallel to the orientifold independently from perpendicular directions; i.e. $(V_1, V_2) = (V_\parallel, V_\perp)$ with $(k_1, k_2) = (1,5)$.

Consider now the scaling behavior of different contributions to the effective potential with the new moduli
\begin{align}
  \label{eq:contributions1}
  \text{Internal Curvature:} \quad &V_\mathcal{R} \propto \tau^{-2}\sigma^{-1},\\
  \text{Orientifold Plane:} \quad &V_O \propto -\tau^{-3}\pi^{1/4}\sigma^{-5/4},\\
  \text{2-form Flux ($\perp$):} \quad &V_2 \propto \tau^{-4}\pi^{1/2}\sigma^{1/2},\\
  \text{2-form Flux ($\parallel$):} \quad &V_2 \propto \tau^{-4}\pi^{-1/2}\sigma^{3/2},\\
  \text{4-form Flux ($\perp$):} \quad &V_4 \propto \tau^{-4}\pi^{1/2}\sigma^{-3/2},\\
  \text{4-form Flux ($\parallel$):} \quad &V_4 \propto \tau^{-4}\pi^{-1/2}\sigma^{-1/2}.
  \label{eq:contributions2}
\end{align}

We take the sum of the above contributions with arbitrary coefficients to be the form of the potential. The existence of a de Sitter minimum requires that the derivative with respect to each modulus vanishes. This condition gives three equations
\begin{align}
- 4A_{4\perp}\pi - 4A_{4\parallel}\sigma - 4A_{2\perp}\pi\sigma^2 - 4A_{2\parallel}\sigma^3 + 3A_O \pi^{3/4}\sigma^{1/4}\tau - 2A_\mathcal{R}\pi^{1/2}\sigma^{1/2}\tau^2 &= 0 \label{eq:deriv1},\\
- 6A_{4\perp}\pi - 2A_{4\parallel}\sigma + 2A_{2\perp}\pi\sigma^2 + 6A_{2\parallel}\sigma^3 + 5A_O \pi^{3/4}\sigma^{1/4}\tau - 4A_\mathcal{R}\pi^{1/2}\sigma^{1/2}\tau^2 &= 0 \label{eq:deriv2},\\
2A_{4\perp}\pi - 2A_{4\parallel}\sigma + 2A_{2\perp}\pi\sigma^2 - 2A_{2\parallel}\sigma^3 - A_O \pi^{3/4}\sigma^{1/4}\tau &= 0 \label{eq:deriv3}.
\end{align}

Forming the linear combination $2\times\ref{eq:deriv1} - \ref{eq:deriv2} + \ref{eq:deriv3}$ gives
\begin{equation}
  A_{4\parallel}+A_{2\perp}\pi\sigma + 2A_{2\parallel}\sigma^2 = 0,
\end{equation}
\noindent which does not have any positive solutions for $\sigma$. Therefore, the three above equations cannot be simultaneously satisfied with physical solutions for all three moduli and we do not have a de Sitter minimum.

We note that in the absence of $F_2$ flux oriented completely in the perpendicular directions (equivalent to setting $A_{2\perp} = 0$), we can derive a bound on $|\nabla V|/V$. Similar to Appendix~\ref{app:typeIIextensions}, we define the derivative $\mathcal{D}(a,b) = -a\tau\partial_\tau - b\sigma\partial_\sigma -\pi\partial_\pi$ and find a tuple $(a,b)$ such that $\mathcal{D}V \geq K V$ with $K>0$. The tuple $(a,b)$ is found by applying the inequalities $C_i^{(-)} \leq \min_j C_j^{(+)}$ where $C_i$ are defined analogously to Appendix~\ref{app:typeIIextensions} but with respect to the contributions \ref{eq:contributions1}-\ref{eq:contributions2}. Minimizing the slow-roll parameter subject to the constraint $\mathcal{D}V/V \geq K$ gives the bound $\epsilon > 1/2$. This bound is not necessarily realized since we do not have a full string theory construction.

\section{Strong and Null Energy Conditions}
\label{app:SECNEC}

In this appendix we will show by a simple generalization of Maldacena-Nu\~{n}ez~\cite{Maldacena:2000mw} and Steinhardt-Wesley~\cite{Steinhardt:2008nk} no-go theorems that one can place a bound on the gradient of the potential. We make the same assumptions as they do (no $\alpha',g_s$-corrections, no D-branes/O-planes, no singularities, all matter satisfies strong/null energy condition. In the Steinhardt-Wesley analysis,  an additional assumption is made
on the geometry of the internal manifold).

The main idea behind the computation is to use $D$-dimensional Einstein equation and strong/null energy condition to get a lower bound on the acceleration of scalar fields in terms of Hubble constant. From the $d$-dimensional point of view, this acceleration is related to the rolling of scalar field down to the lower values of the potential energy. Since the value of the acceleration is related to the gradient of the potential, the lower bound on the acceleration is related to a lower bound on the value of the gradient.

For example, the Maldacena-Nunez argument can be understood as follows. Let us assume that there exists a stationary point $V'(\Phi_0)=0$ with positive energy density $V(\Phi)>0$. Then, we can solve the effective $4d$ equations of motion and find a stationary de Sitter solution of supergravity. However, a simple manipulation with the
$10d$ Einstein equations shows that this solution is incompatible with the strong energy condition. Therefore, we conclude that the effective potential does not have stationary points.

We will make one step further and use this argument to place a lower bound on the gradient of the potential. We will place our system at any point of the potential with positive energy $V(\Phi)>0$ and we will choose the initial condition to be $\partial_t\Phi=0$, $i.e.$ speed of all fields is set to zero. We can easily solve the effective $d$-dimensional equations
of motion and find that our field will have a non-zero acceleration, which is related to the gradient of the potential energy. On the other hand, we can use the $D$-dimensional Einstein equation of motion in the same way as in Maldacena-Nunez paper to get a lower bound on the acceleration. Combining it with the effective $d$-dimensional point of view, we will find our bound on the gradient of the potential.

\subsection{Strong Energy Condition}
Consider a $D$-dimensional metric with $d$ macroscopic dimensions
of the form
\begin{equation}
ds^2= \Omega(y,t)^2\left( -dt^2 + a(t)^2 d\vec x^2_{d-1}\right) +g_{mn}(y,t) dy^m dy^n.
\end{equation}
We use capital letters $M,N,\dots$ for $D$ dimensional indices, Greek letters $\alpha,\beta,\dots$ for $(d-1)$-dimensional space, $0$ for time, and small Latin letters for compact directions. We assume that the warp factor $\Omega$ and internal metric $ g_{mn}$ depend only on the internal coordinates $y$ and time $t$.

This ansatz has a residual diffeomorphism symmetry related to the definition of our $d$-dim\-ensional frame $t,\vec x$. We will fix this symmetry by choosing the $d$-dimensional gravitational coupling $\kappa_d$ to be equal to $D$-dimensional one $\kappa_D$. For warped volume this means that
\begin{align}\label{frame choice}
\int dy^{D-d}\,\Omega^{d-2}\sqrt{\det  g_{mn}}=1.
\end{align}
This in particular implies
\begin{align}\label{gauge choice}
\int dy^{D-d}\,\Omega^{d-2}\sqrt{\det  g_{mn}}\left((d-2)\frac{\partial_{0}^2\Omega}{\Omega} +\frac 1 2  g^{mn}\partial_{0}^2 g_{mn}\right) = 0.
\end{align}

We will tune initial conditions to make $\dot \Omega \equiv \partial_t \Omega = 0$ and $\dot g_{mn} \equiv \partial_t  g_{mn}  = 0$ at some moment in time, say $t=0$. All quantities will be assumed to be computed at this moment in the following. From the $D$-dimensional perspective, this is a choice of the initial conditions for the metric
in solving the Einstein equations. From the $d$-dimensional perspective, this is a choice of the initial condition $\dot \Phi=0$ for all moduli fields $\Phi$. Note, that we cannot set acceleration $\ddot\Phi$ to zero since it will be incompatible with the equations of motion. From the $d$-dimensional perspective, we have scalar fields with potential energy $V(\Phi)$ and no kinetic energy. Therefore, $d$-dimensional FRW equations give ${\ddot a}/a={\dot a}^2/a^2=H^2$ at $t=0$ where the Hubble constant $H$ is determined by the value of $V$.

Straightforward computation of the  $00$-component of Ricci curvature gives
\begin{align}
R_{00}=-\frac 1 2  g^{mn}\ddot g_{mn} -(d-1)\frac{\ddot a}a-(d-1)\frac{\ddot \Omega}{\Omega}+ \frac{1}{d}\Omega^{2-d}\nabla_m\nabla^m  \Omega^d,
\end{align}
where  $ \nabla_m$ is $(D-d)$-connection computed with respect to the metric $ g_{mn}$.

We want to find a restriction on the values of the acceleration using the strong energy condition in $D$-dimensions. It gives
\begin{align}\label{tr einst eq}
\begin{split}
&0\le T_{00}+\frac{1}{D-2} T^N_N = R_{00}  =-\frac 1 2  g^{mn}\ddot g_{mn} -(d-1)\frac{\ddot a}a-(d-1)\frac{\ddot \Omega}{\Omega}+ \frac{1}{d}\Omega^{2-d}\nabla_m\nabla^m  \Omega^d,
\end{split}
\end{align}
where the first inequality is a definition of the strong energy condition in $D$-dimensions and the second equality follows from the $D$-dimensional Einstein equations. Multiplying (\ref{tr einst eq}) by $\Omega^{d-2}$, integrating over the compact manifold and using $\ddot a/a=H^2$, we get
\begin{align}\label{SEC bound}
(d-1)H^2  \le -  \int dy^{D-d}\,\Omega^{d-2}\sqrt{\det g_{mn}}\, \frac{\ddot\Omega}{\Omega},
\end{align}
where we have used the condition (\ref{gauge choice}) and the fact that the integral of the Laplacian of the warp factor over the whole manifold is zero provided the manifold has no boundary and the warp factor is non-singular.

In this way, using the $D$-dimensional Einstein equations, we found a lower bound on the second derivative of $\Omega$ which is related to the acceleration of the scalar fields. In order to relate this restriction to a bound on the potential, we need to analyze the $d$-dimensional effective action carefully.

Note, that the inequality gives a bound on the value of the scalar field acceleration only in the direction of the overall volume change. Other deformations of the manifold are unimportant. So we can define a basis in such a way that one mode, which we denote $\tau$, is the overall volume modulus, and other fields are chosen to be orthogonal to it (in a sense that only $\ddot \tau$ gives a nonzero contribution to the right-hand side of (\ref{SEC bound})).
To relate the bound on the acceleration to a bound on the derivative of the potential we need to find the $d$-dimensional effective action. As was noted above, only overall scaling mode $\tau$ matters. Separating this mode from the others,
we write,
\begin{equation}
ds^2= e^{-2\sqrt{\frac{D-d}{(D-2)(d-2)}}\tau} \widetilde{\Omega}(y,t)^2 \left( -dt^2 + a(t)^2 d \vec x^2_{d-1}\right) + e^{2\sqrt{\frac{d-2}{(D-d)(D-2)}}\tau} \widetilde{g_{mn}}(y,\Phi) dy^m dy^n,
\end{equation}
where tilde indicates that the overall modulus contribution was singled out.
The relevant part of the effective action is
\begin{align}
\begin{split}
S= \int dx^d \sqrt{\det g_d} \left(\frac{1}{2\kappa_d^2}R_d+\frac{1}{2\kappa_d^2}\dot \tau^2-V(\tau,\Phi)+\text{other fields contribution}\right),
\end{split}
\end{align}
where $\kappa_d$ is the $d$-dimensional gravitational coupling constant and $\Phi$ stands for all other fields.

The $d$-dimensional Einstein equations and the equation of motion for $\tau$ are (at the point, where $\dot \tau=0$ and $\dot \Phi=0$ for all other fields),
\begin{align}
H^2=\frac{2\kappa_d^2   }{(d-1)(d-2)}V(\tau,\Phi),\\
\ddot\tau+\kappa_d^2\,\partial_{\tau} V(\tau,\Phi)=0.
\end{align}
Combining these equations with the SEC bound (\ref{SEC bound}), we obtain the following
bound on the gradient of the potential,
\begin{align}\label{SEC bound result}
-\frac{\partial_{\tau} V(\tau,\Phi)}{V(\tau,\Phi)}\ge 2\sqrt{\frac{D-2}{(D-d)(d-2)}}=\lambda_{SEC}
\end{align}

\subsection{Null Energy Condition}

The strong energy condition is known to be violated in string theory and it is natural to try to relax it.
We have been able to do so only with an additional assumption: we will follow
 \cite{Steinhardt:2008nk} to impose a restriction on the geometry of the internal manifold. The restriction
 -- to be specified later -- can be conveniently expressed in terms of a conformally transformed metric $g_{mn}(y,t)$,
\begin{equation}
ds^2= \Omega(y,t)^2\left( -dt^2 + a(t)^2 d\vec x^2_{d-1}\right) +\Omega(y,t)^{-\frac{2d}{D-d-2}} g_{mn}(y,t) dy^m dy^n,
  \label{NECansatz appendix}
\end{equation}
We chose the power of $\Omega$ for two reasons. First, it will simplify the assumption
on the internal metric $g_{mn}(y,t)$. Secondly, this ansatz is compatible with many explicit solutions in string and M-theory,
such as the KKLT construction, brane solutions, and M/string theory flux compactifications.

After tuning initial conditions $\dot \Omega = \dot g_{mn}=0$  and fixing the frame,
\begin{align}\label{frame fix NEC}
\int dy^{D-d}\,\Omega^{-2\frac{D-2}{D-d-2}}\sqrt{\det  g_{mn}}=1.
\end{align}
the relevant components of the Ricci tensor are expressed as,
\begin{align}
R_{00}=-\frac 1 2 g^{mn}\ddot g_{mn}-(d-1)\frac{\ddot a}a+\frac{D+d-2}{D-d-2}\frac{\ddot \Omega}{\Omega} + \Omega^{2\frac{D-2}{D-d-2}}\nabla_m\nabla^m \log \Omega,
\end{align}
\begin{align}
\begin{split}
R_{mn}= R_{mn}^{(D-d)}+\frac{1}{2} \Omega^{-2\frac{D-2}{D-d-2}}\ddot g_{mn} -\frac{d}{D-d-2} \Omega^{-2\frac{D-2}{D-d-2}}\frac{\ddot  \Omega}{\Omega} g_{mn}\\+\frac{d}{D-d-2} g_{mn}  \nabla^2 \log \Omega -\frac{d(D-2)}{D-d-2} \nabla_m \log\Omega\nabla_n \log\Omega,
\end{split}
\end{align}
where  $ R_{mn}^{(D-d)}$, $ \nabla_m$ are $(D-d)$-dimensional quantities computed with respect to the metric $ g_{mn}$.

Using the null energy condition for a null vector $n^N$ with non-zero components in time and compact directions, we get
\begin{align}\label{nec cond}
\begin{split}
0\le T_{MN}n^M n^N = R_{MN} n^M n^N,
\end{split}
\end{align}
where the first inequality is the definition of the null energy condition and the second follows from Einstein equation. We choose a null vector with zero components in the space directions, and average it over the compact direction. The result is $\langle n^0 n^0\rangle= \Omega^{-2}$  and $\langle n^m n^n\rangle= \frac 1 {D-d}\Omega^{\frac{2d}{D-d-2}} g^{mn}$, where $\langle\, \dots\rangle$ means averaging. After averaging, equation  (\ref{nec cond}) takes the form
\begin{align}\label{tr einst eq NEC}
\begin{split}
0\le T_{MN}\langle n^M n^N\rangle = R_{MN}\langle n^M n^N \rangle=\frac{1}{(D-d)}\Omega^{\frac{2d}{D-d-2}} R^{(D-d)}-\frac{d-1}{\Omega^2} \frac{\ddot a}a  -\frac{D-d-1}{2(D-d)\Omega^2} g^{mn}\ddot g_{mn}\\+\frac{D-2}{D-d-2}\frac{\ddot \Omega}{\Omega^3} +\frac{D-2}{D-d-2}\Omega^{\frac{2d}{D-d-2}} \nabla_m\nabla^m \log\Omega- \frac{d(D-2)}{(D-d-2)(D-d)}\Omega^{\frac{2d}{D-d-2}} \left( \nabla \log\Omega\right)^2,
\end{split}
\end{align}
where $ R^{(D-d)}$ is scalar curvature of the metric $ g_{mn}$.

Multiplying (\ref{tr einst eq NEC}) by $\Omega^{\frac{-2d}{D-d-2}}$, integrating over the compact manifold and using $\ddot a/ a =H^2$, we find,
\begin{align}\label{nec ineq}
\begin{split}
(d-1)H^2  &\le\,\,  \int dy^{D-d}\,\sqrt{\det  g_{mn}}\Big[ -\frac{D-2}{D-d} \Omega^{-2\frac{D-2}{D-d-2}}\frac{\ddot\Omega}{\Omega}\\
&+\frac{1}{(D-d)} R^{(D-d)}- \frac{d(D-2)}{(D-d-2)(D-d)} \left( \nabla \log\Omega\right)^2\Big],
\end{split}
\end{align}
where we have used condition (\ref{frame fix NEC}) and that the integral of the Laplacian of the warp factor over the whole manifold is zero.

The assumption we make on the metric $g_{mn}$ for the internal manifold is the following:
We assume that the integral over the compact manifold of the scalar curvature is either zero or negative.
This condition is motivated by explicit examples such as KKLT, where the metric takes the form (\ref{NECansatz appendix}) with the internal metric $g_{mn}$ being Ricci flat.  We also assume $D \geq d+2$ so that
the last coefficient in the right-hand side of (\ref{nec ineq}) is non-negative.
The ansatz becomes singular when $D=d+2$, but this singularity is not important.
After the redefinition $\Omega \rightarrow \Omega^{(D-d-2)}$  similar computations lead
to the same bound (\ref{NEC bound result}).

In this case, the following bound holds
\begin{align}\label{NEC bound}
(d-1)H^2  \le -\int dy^{D-d}\,\Omega^{-2\frac{D-2}{D-d-2}}\sqrt{\det  g_{mn}} \frac{D-2}{D-d}\frac{\ddot\Omega}{\Omega}.
\end{align}
Again, using the $D$-dimensional Einstein equations we found a lower bound on the acceleration of the scalar fields. Following the same steps as in SEC case, we get the following bound on the gradient of the potential
\begin{align}\label{NEC bound result}
-\frac{\partial_{ {\tau}} V(\Phi)}{V(\Phi)}\ge 2\sqrt{\frac{D-d}{(D-2)(d-2)}}= \lambda_{NEC}.
\end{align}
\bibliographystyle{JHEP}
\bibliography{references}

\end{document}